# Magnetic and magnetoresistance behavior of $Tb_7Rh_3$, an intermetallic compound with a negative temperature coefficient of electrical resistivity in the paramagnetic state, and "Paramagnetic Giant Magnetoresistance Phenomenon"

Kausik Sengupta, Kartik K Iyer and E.V. Sampathkumaran[*]

*Tata Institute of Fundamental Research, Homi Bhabha Road, Mumbai-400005, India*

The results of dc magnetization, electrical and magnetoresistance and heat capacity measurements (2-300 K) on $Tb_7Rh_3$, crystallizing in $Th_7Fe_3$-type hexagonal structure, are reported. In this compound, magnetic ordering sets in around 90 K with additional transitions at low temperatures and the temperature coefficient of resistivity ($\rho$), $d\rho/dT$, is negative over a wide temperature range in the paramagnetic state. The present magnetization results reveal that this compound is apparently characterized by rich features in the magnetic-field-temperature phase diagram. A point of major emphasis is that the sign of $d\rho/dT$ in the paramagnetic state can be gradually changed by the application of magnetic field. As a result, the magnitude of the magnetoresistance (MR) is rather large even in the vicinity of room temperature (far above magnetic ordering temperature), in addition to giant MR behavior in the magnetically ordered state. Viewed together With similar behavior for other heavy rare-earth members of this series, this class of compounds can be classified as "Paramagnetic Giant Magnetoresistance Systems". A new theoretical approach is warranted to understand this phenomenon.

PACS numbers:

| | |
|---|---|
| 75.47.De | Giant magnetoresistance |
| 72.15.-v | Electronic conduction in metals and alloys |
| 75.50.-y | Studies of specific magnetic materials |

[*]E-mail: sampath@tifr.res.in

Among the rare-earth (R) intermetallics, the binary compounds were extensively studied during last four decades. However, surprisingly, a binary series containing Rh, namely $R_7Rh_3$ (Ref. 1), was not subjected to much investigations, barring some reports by Tsuotaka et al [2,3] on the magnetization (M) and electrical resistivity ($\rho$) behavior. These results revealed that the $\rho(T)$ behavior of these compounds is quite exceptional among R-intermetallics in the sense that the temperature (T) coefficient of $\rho$ for the heavy R members is negative in the paramagnetic state. This anomaly was believed to arise from semiconducting gap effects and was also proposed to be sensitive to lanthanide contraction considering that, for light R members, $d\rho/dT$ is positive. In recent years, we have reported the results of magnetoresistance (MR) studies for R= Gd, Dy and Er [4] and we have argued that the paramagnetic transport anomaly originates from an unusual spin-disorder contribution, and not due to a semiconducting gap; sign of $d\rho/dT$ could be made positive by the application of a magnetic field (H) as a result of which we observed unusually large MR even near 300 K in the paramagnetic state. With the primary motivation of bringing out that this is a general characteristic behavior among heavy R members of this class of compounds, we have carried out detailed investigations on the Tb case, for which the onset of antiferromagnetic ordering has been shown to take place near 90 K on the basis of M and $\rho$ measurements not only on polycrystals but also on single crystals [3].

The polycrystalline sample, $Tb_7Rh_3$, was prepared by arc melting stoichiometric amounts of Tb and Rh. The molten ingot was annealed at 600 C for 48 hrs. The sample thus obtained was found to be a single phase by x-ray diffraction. The dc magnetic susceptibility ($\chi$) measurements in a magnetic-field (H) of 5 kOe (2-300 K) and 100 Oe (2 – 120 K) and isothermal M measurements at selected temperatures (up to 120 kOe) were performed by a commercial (Oxford Instruments) vibrating sample magnetometer. The $\rho$ measurements (2-300 K) were performed up to 140 kOe employing the Physical Property Measurements System (Quantum Design) and the same commercial instrument was employed to collect the heat-capacity (C) data.

In figure 1 (top), we show the behavior of $\chi(T)$, obtained for the zero-field-cooled (ZFC) condition of the specimen. $\chi$ follows Curie-Weiss behavior above 90 K. The effective moment obtained from the linear region is about 11.8 $\mu_B$/Tb, which is nearly 2 $\mu_B$ higher than that expected for free trivalent Tb ion and this is attributed to the polarization induced in the Rh 4d band. This value is about 1 $\mu_B$ is larger than that reported in Ref. 3, but the magnitude of the paramagnetic Curie temperature is much smaller (-4 K) than the reported value (-45 K). As reported in Ref. 3, there is a sharp kink at 90 K indicative of the onset of an antiferromagnetic transition. As the T is lowered further, we observe a $\chi(T)$ feature, which is different from that in Ref. 3 for polycrystals. There is a pronounced upturn below about 75 K, the temperature at which a ferrimagnetic ordering was proposed; nearly at the same temperature, $\chi(T)$ curves, obtained for the ZFC and field-cooled (FC) conditions of the specimen with H= 100 Oe, bifurcate (Fig. 1 bottom), as though there are dramatic changes in magnetism around this temperature. This is followed by a peak around 38 K and a fall at lower temperatures in the ZFC curves. There is a sharp fall of $\chi$ near 15 K for H= 5 kOe, whereas in Ref. 3, a flattening of the plot is seen as though, in the latter case, the condition of measurements corresponds to field-cooling of the sample. Possibly strong orientation effects of the crystallites also may be responsible for the discrepancies between the two reports on polycrystals. Despite these discrepancies, our data support the conclusion broadly that there are additional magnetic transitions and it is not clear whether these transitions arise from three crystallographically inequivalent Tb ions or whether there are temperature-dependent spin-reorientation effects. In order to bring out that the magnetic structure could be different in various temperature ranges in the magnetically ordered state, we show the isothermal M behavior at selected temperature ranges in figure 2. At 5 K, the zero-field state appears to be antiferromagnetic as indicated by a step around 5 kOe (shown in the inset of figure 2 (top)) followed by a linear H-dependence of M till about 70 kOe in the forward cycle of the M-H curve. In addition, there is a field-induced antiferro-to-ferromagnetic transition at high fields near 100 kOe [3], which is hysteretic in nature, thereby establishing first-order nature of the transition. In contrast to the observation on single crystals [3], we see a small irreversibility (or a loop) below 15 kOe at this temperature, which indicates that a ferromagnetic component is present at 5 K after the field cycling. Ref. 3 does not report isothermal M data at very high fields at other temperatures. In order to demonstrate how M-H plots look as the T is increased, we now look at the data at 30 and 50 K in the figure 2. In contrast to Ref. 3, we do not see hysteresis at low fields. In addition, we notice that there are additional sharp steps in the very high field region, however, with a weaker hysteresis compared to 5 K. These findings establish that the nature of antiferromagnetic structure varies with temperature. It is therefore of interest to carry out neutron diffraction studies on this compound. These steps vanish (as expected) in the paramagnetic state as demonstrated by the M(H) plots at 100, 120 and 150 K in figure 2. We have also carried out heat capacity measurements in order to probe the magnetic transitions (see Fig. 3). While there is a distinct peak at 91 K confirming the onset of magnetic order, no additional peak could be observed in the data at lower temperatures, particularly near 15-25 or 70 K (at which $\chi$ and $\rho$ data reveal the existence of additional transitions). This implies that the entropy change involved at the low temperature transitions are rather small. C varies as $T^3$ below 15 K as shown in the inset of figure 3, typical of antiferromagnets, supporting the inferences from M data.

*We now present the transport behavior, which is the point of central emphasis. The $\rho(T)$ behavior in the presence of various external H are shown in figure 4.*

We first focus on the paramagnetic behavior. The sign of d$\rho$/dT is negative as in Ref. 3 till magnetic ordering sets in. The application of H tends to change this sign gradually as inferred from the plots in figure 3; for H= 80 kOe, a broad maximum could be observed around 150 K, resembling the zero-field behavior of $Dy_7Rh_3$ [3, 4], with a noticeable depression of absolute values of $\rho$. With a further application of H, say 140 kOe, the temperature coefficient of $\rho$ becomes positive, resulting in a significant MR. In order to have a better idea of the magnitude of MR, we have also measured $\rho$ as a function of H at fixed temperatures (see figure 5). At 300 K, the sign of MR [defined as ($\rho$(H)- $\rho$(0))/$\rho$(0)] is positive for initial applications of H, which implies that the Lorentz contribution is the dominant one; for higher values of H, spin disorder contribution dominates, as indicated by the negative sign of MR. At 250 K, the sign of MR is distinctly negative for all H and MR varies quadratically with H typical of paramagnets. Interestingly, as in the case of Gd and Dy analogues [4], the magnitude of MR is quite large far above Néel temperature, even near room temperature, e,g., about -3.5% at 250 K and about -10% at 150 K for H=140 K. This property is uncharacteristic of paramagnets. These findings naturally imply that the transport anomaly arises from an unusual spin-disorder contribution [4].

Turning to the magnetically ordered state, the nature of the $\rho(T)$ plot is essentially the same as in Ref. 3. There is an upturn below 90 K due to the formation of magnetic Brillouin-zone boundary gaps; there is a kink near 70 K which is clearly revealed by the existence a peak in *d$\rho$/dT* in the inset of figure 3. It appears that the magnetic structure modification occurring around 70 K changes the Fermi surface characteristics, as a result of which the sign of d$\rho$/dT becomes positive below this temperature. Initial applications of H till 80 kOe influences the $\rho$ behavior in a narrow range of temperature only (70 to 90 K) below the onset of magnetic transition, without causing much effect below 70 K. This MR behavior support the idea that the magnetic states below and above 70 K are different. If a very large field beyond metamagnetic transitions, e.g., 140 kOe, is applied, the $\rho$ values are dramatically reduced even below 70 K, resulting in positive d$\rho$/dT in the entire T-range of investigation. Clearly, there is a large MR for such fields. In order to bring out this point, we show the MR derived from the $\rho(T)$ plots in figure 4. There is a well-defined peak for H= 50 and 80 kOe with a maximum around 90 K, whereas for H= 140 kOe, MR keeps increasing with decreasing temperature. We also show MR as a function of H in figure 5 at 80 and 100 K. We are not able to get reliable data at lower temperatures at very high fields due to the rotation of the sample by the torque following metamagnetic transitions and at lower fields MR values are negligible small as inferred from figure 4 (top) (and hence not shown in figure 5) . The magnitudes of MR shown in figures 4 and 5 implies that this compound exhibits giant magnetoresistance (GMR) behavior in the magnetically ordered state as well.

Summarizing, the binary intermetallic compound, $Tb_7Rh_3$, is established to be an exotic one with a negative temperature coefficient of electrical resistivity in the paramagnetic state over a wide temperature range [3]. The magnetization results indicate that this compound could be characterized by an interesting phase diagram in the H-T plane and it is worthwhile to carry out neutron diffraction studies in this regard. An application of magnetic field can restore 'metallic' $\rho(T)$ behavior as though there is a field-induced semiconductor-to-metal transition, resulting in large magnetoresistance in the paramagnetic state near room temperature, apart from GMR behavior in the magnetically ordered state, like R=Gd and Dy members of this series. Therefore, these compounds can be classified as "paramagnetic GMR systems". Needless to reiterate [4] that a theory is required to understand this transport anomaly in the paramagnetic state of antiferromagnetic systems, though there were some recent attempts for ferromagnetic systems [5]. There is also an urgent need to explore whether the transport anomaly for heavy rare-earth members is caused by the Kondo effect (or any other narrow-band phenomenon) induced in the 4d band of Rh; if so, this series provides an unique opportunity to open up investigations of such phenomena among concentrated 4d systems. Finally, we would like to add that $Gd_7Rh_3$ is found to exhibit large magnetocaloric effects at the magnetic transition (140 K) [6], whereas for the Tb analogue, this effect is found to be negligible as inferred from isothermal M data measured at close intervals of temperature.

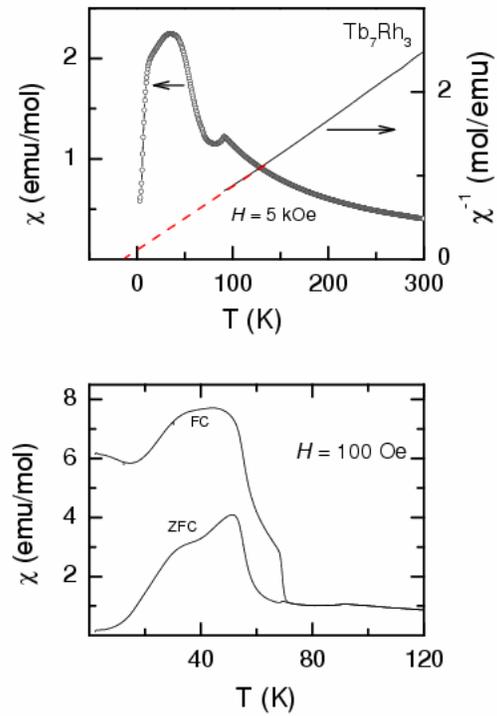

Figure 1:

(Top) Magnetic susceptibility ($\chi$) as a function of temperature for $Tb_7Rh_3$, measured in a field of 5 kOe for the zero-cooled-cooled condition of the specimen. The plot of inverse $\chi$ is also shown in the paramagnetic state and the dashed line represents linear extrapolation to the magnetically ordered state. (Bottom) The data recorded for the ZFC and field-cooled (FC) conditions of the specimen for H= 100 Oe.

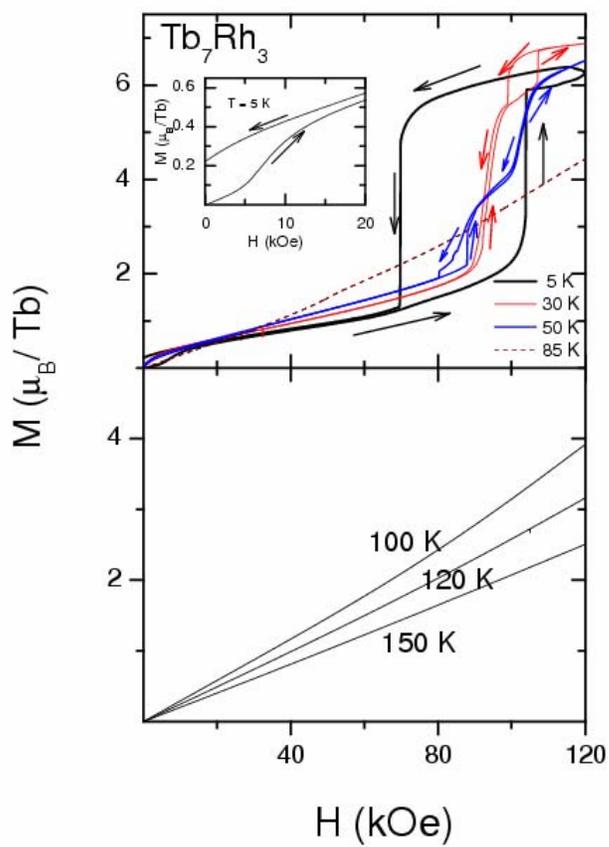

Figure 2:

Isothermal magnetization behavior for $Tb_7Rh_3$ at selected temperatures (5, 30, 50, 85, 100, 120 and 150 K) for the zero-field-cooled condition of the specimen. The arrows show the direction of variation of magnetic field. In the inset of top figure, the data for 5 K below 20 kOe is shown in an expanded form to highlight the presence of a step in the forward cycle in the virgin curve.

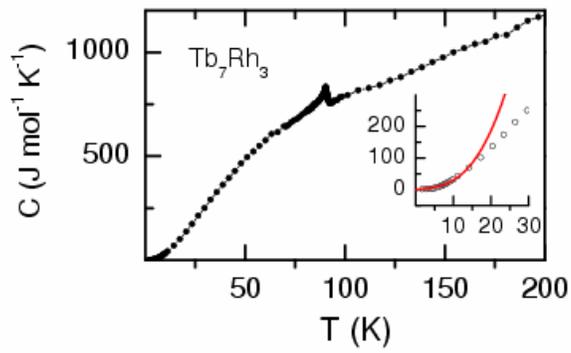

Figure 3:
Heat capacity as a function of temperature for $Tb_7Rh_3$. The data below 30 K is plotted in the inset to highlight $T^3$ dependence below 15 K.

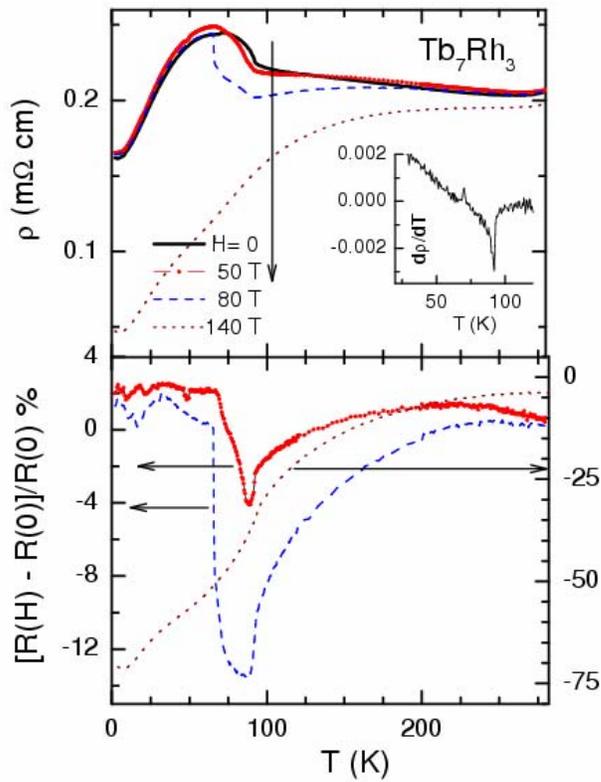

Figure 4:

Electrical resistivity (ρ) as a function of temperature for Tb$_7$Rh$_3$ in the presence of various magnetic fields. The existence of a transition around 70 K is more clearly seen in the plot of dρ/dT shown in the inset. The magnetoresistance obtained from this data is also plotted in the bottom figure.

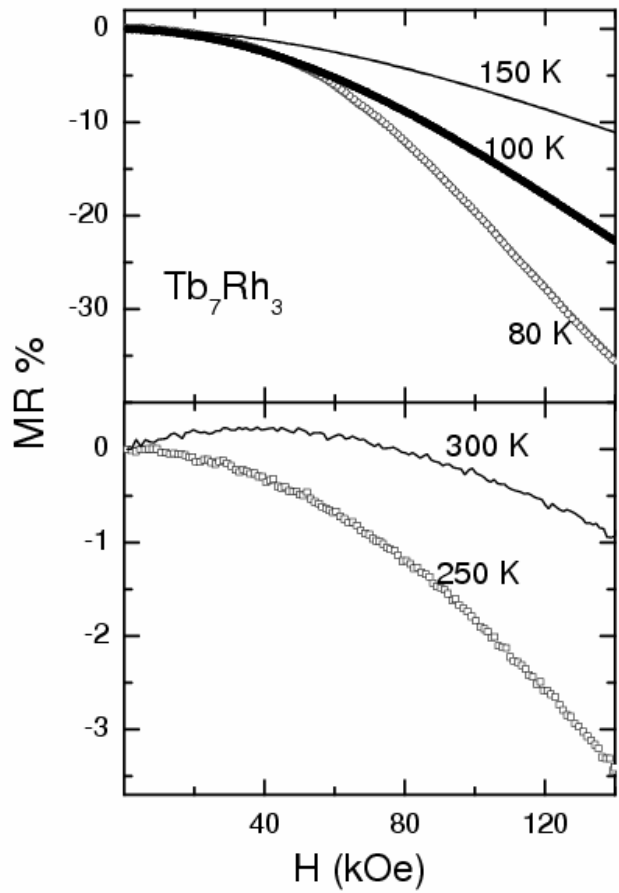

Figure 5:

Magnetoresistance as a function of H for $Tb_7Rh_3$ at selected temperatures (plot for 80, 100, 150, 250 and 300 K).